\documentclass{SciPost}

\hypersetup{
    colorlinks,
    linkcolor={red!50!black},
    citecolor={blue!50!black},
    urlcolor={blue!80!black}
}

\usepackage[bitstream-charter]{mathdesign}
\DeclareSymbolFont{usualmathcal}{OMS}{cmsy}{m}{n}
\DeclareSymbolFontAlphabet{\mathcal}{usualmathcal}

\fancypagestyle{SPstyle}{
\fancyhf{}
\lhead{\colorbox{scipostdeepblue}{\bf \color{white} ~SciPost Physics Proceedings }}
\rhead{{\bf \color{scipostdeepblue} ~Submission }}

\fancyfoot[C]{\textbf{\thepage}}
}

\begin{document}

\pagestyle{SPstyle}

\begin{center}{\Large \textbf{\color{scipostdeepblue}{
Domain-Specific Agents for Cherenkov Telescope Array Control Software and Gamma-Ray Data Analysis
}}}\end{center}

\begin{center}\textbf{
Dmitriy~Kostunin\textsuperscript{1$\star$},
Elisa~Jones\textsuperscript{1},
Vladimir~Sotnikov\textsuperscript{2},
Valery~Sotnikov\textsuperscript{3},
Sergo~Golovachev\textsuperscript{3} and
Alexandre Strube\textsuperscript{4}
}\end{center}

\begin{center}
{\bf 1} Deutsches Elektronen-Synchrotron DESY, 15738 Zeuthen, Germany
\\
{\bf 2} JetBrains Limited, 8046 Paphos, Cyprus
\\
{\bf 3} JetBrains GmbH, 80639 München, Germany
\\
{\bf 4} Jülich Supercomputing Centre, 52425 Jülich, Germany
\\
[\baselineskip]
$\star$ \href{mailto:dmitriy.kostunin@desy.de}{\small dmitriy.kostunin@desy.de}
\end{center}

\definecolor{palegray}{gray}{0.95}
\begin{center}
\colorbox{palegray}{
  \begin{tabular}{rr}
  \begin{minipage}{0.37\textwidth}
    \includegraphics[width=60mm]{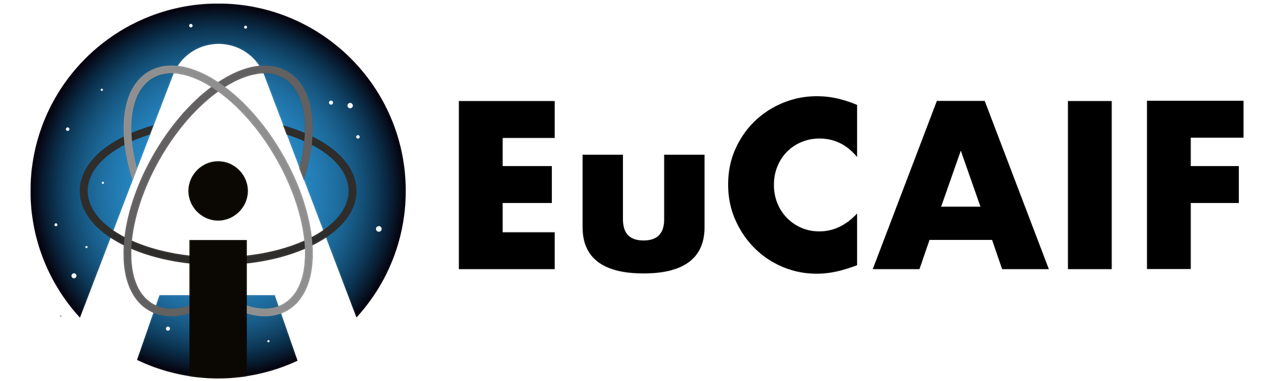}
  \end{minipage}
  &
  \begin{minipage}{0.5\textwidth}
    \vspace{5pt}
    \vspace{0.5\baselineskip} 
    \begin{center} \hspace{5pt}
    {\it The 2nd European AI for Fundamental \\Physics Conference (EuCAIFCon2025)} \\
    {\it Cagliari, Sardinia, 16-20 June 2025
    }
    \vspace{0.5\baselineskip} 
    \vspace{5pt}
    \end{center}
    
  \end{minipage}
\end{tabular}
}
\end{center}

\section*{\color{scipostdeepblue}{Abstract}}
\textbf{\boldmath{%
We present domain-adapted large language model agents designed to support Cherenkov Telescope Array operation and data analysis. The agents combine contextual knowledge with automated validation and iterative correction to produce more reliable outputs. This approach reduces manual effort, improves consistency, and helps accelerate operational and scientific workflows. The results demonstrate the potential of agentic systems as practical assistants in specialized research environments.
}}

\vspace{\baselineskip}

\noindent\textcolor{white!90!black}{%
\fbox{\parbox{0.975\linewidth}{%
\textcolor{white!40!black}{\begin{tabular}{lr}%
  \begin{minipage}{0.6\textwidth}%
    {\small Copyright attribution to authors. \newline
    This work is a submission to SciPost Phys. Proc. \newline
    License information to appear upon publication. \newline
    Publication information to appear upon publication.}
  \end{minipage} & \begin{minipage}{0.4\textwidth}
    {\small Received Date \newline Accepted Date \newline Published Date}%
  \end{minipage}
\end{tabular}}
}}
}
\vspace{10pt}
\noindent\rule{\textwidth}{1pt}
\tableofcontents
\noindent\rule{\textwidth}{1pt}
\vspace{10pt}

\section{Introduction}
LLMs excel at generating code, but specialized scientific domains pose challenges: unusual data formats, fast-changing APIs, and strict reproducibility, so off-the-shelf help must be reinforced with expert knowledge and guardrails. General models trained on broad code often falter with rapidly evolving or poorly documented frameworks common in cutting-edge science.

The upcoming Cherenkov Telescope Array Observatory (CTAO)\footnote{\url{https://www.ctao.org/}} exemplifies this challenge. Its Array Control and Data Acquisition software (ACADA) is a heterogeneous stack orchestrating dozens of telescopes and instruments. Developing and maintaining ACADA demands extensive domain expertise, careful adherence to project conventions, and repetitive boilerplate coding~\cite{Oya:2024dww}. At the other end of the workflow, scientists will analyze CTAO data using open-source tools like Gammapy~\cite{gammapy:2023}, a Python library that provides a common platform to reduce and model telescope observations. While Gammapy is powerful, its specialized focus and evolving API can make it difficult for a generic LLM (trained on limited or outdated data) to produce correct analysis scripts without additional guidance~\cite{2025arXiv250300821K}. 

\begin{figure}[h]
    \centering
    \includegraphics[width=0.95\textwidth]{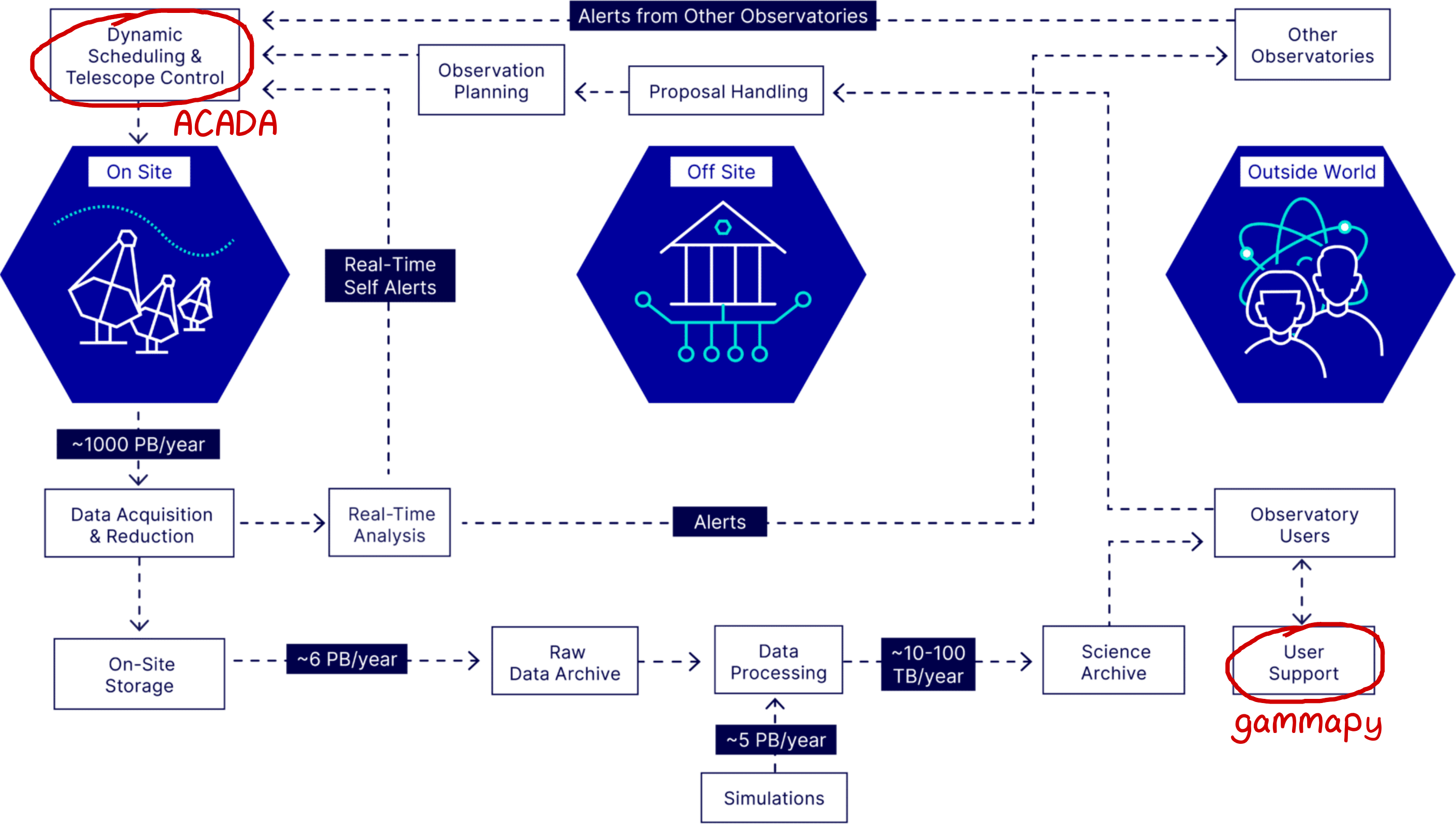}
    \caption{CTAO data flow highlighting integration points for the LLM-based assistants. CTAgent operates at the level of observatory operations and control software (left). Gammapygpt operates at the scientific analysis stage (right), helping researchers to produce and validate analysis scripts based on the Gammapy library.}
    \label{fig:ctaflow}
\end{figure}

We present two LLM-driven agents designed to assist the CTAO workflow from operations to analysis. CTAgent~\cite{Kostunin:2025rlk} is an agent tailored for the development of ACADA control software; it ingests CTAO design artifacts and synthesizes validated code to accelerate the implementation of control systems. Gammapygpt~\cite{2025arXiv250926110} is a complementary agent focused on scientific data analysis; it generates and verifies Gammapy-based analysis scripts for common gamma-ray astronomy tasks. By incorporating project-specific context and automated self-correction, these systems aim to reduce human workload and error rates across the gamma-ray data pipeline. Figure~\ref{fig:ctaflow} illustrates a simplified CTAO data flow and highlights where each agent operates in the process.

\section{CTAgent}
CTAgent is being developed to reduce the manual effort involved in writing CTAO control and operations software. The first feature of CTAgent is to automate this process of defining data models and configuration schemas for ACADA. Its design breaks the problem into distinct stages:
\begin{itemize}
    \item \textit{Input ingestion:} All input specification files are first converted to a unified text representation. This preprocessing step ensures that binary or structured documents are normalized for analysis, enabling the agent to handle diverse document formats in a consistent way.
    \item \textit{LLM-based code synthesis:} The agent employs a set of specialized LLM ``experts'' to propose code for a given input. Each expert is tuned (via prompting) to a particular file type or role. 
    \item \textit{Validation and self-repair:} An orchestrator module automatically checks the generated code. It tries to parse the code and applies structural tests. If any check fails the agent enters a refinement loop.
\end{itemize}

CTAgent has been tested on a variety of CTAO use cases, such as generating data structure classes from interface control documents. Even with relatively compact LLMs, the agent was able to produce correct, ready-to-use code for most inputs after a few iterations of self-correction. In essence, CTAgent serves as a pathfinder for introducing AI-assisted development into CTAO operations software, offering insights into where such agentic workflows can add the most value in the software lifecycle.

\section{Gammapygpt}
While CTAgent assists with observatory operations software, Gammapygpt is designed to support the scientific analysis phase by writing and correcting analysis scripts. Its goal is to empower researchers to obtain working Gammapy-based analysis code from a high-level problem description. The agent accepts a natural-language prompt describing a desired analysis (for example, \emph{``Generate a spectral analysis for source Mkn~421 using all available CTAO observations''}), and returns a complete Python script to perform that analysis. Critically, it also verifies that the script runs properly. To accomplish this, agent follows several key principles:

\begin{itemize}
    \item \textit{Structured prompting and strict output format:} The agent’s system prompt imposes firm rules on the LLM’s responses. The model is instructed to produce exactly one self-contained Python code listing, including all necessary import statements and without any extraneous commentary or interactive input prompts.
    \item \textit{Sandboxed execution of code:} Each candidate script is executed in a controlled sandbox environment that mirrors a typical analysis setup. The agent ensures the script knows where to find the data (e.g. by setting an environment variable that points to a mounted dataset directory) and captures any output or errors.
    \item \textit{Iterative error feedback:} If the script fails to run to completion or does not produce the expected result (for example, if it crashes or the output values are out of an acceptable range), the agent automatically analyzes the log and extracts a concise error summary. This summary is then fed back into the LLM on the next prompt iteration (prefixed by a system message indicating the script failed and highlighting the error).
\end{itemize}

Initial benchmarking indicates that Gammapygpt can achieve high reliability on routine Gammapy tasks. In tests comparing OpenAI's o3 and gpt-5, both models eventually reached a 100\% success rate on straightforward analysis tasks, and the more advanced model typically required fewer iterations to fix errors. These outcomes demonstrate that a properly constrained LLM can navigate the relatively unfamiliar domain of Gammapy and produce useful code, as long as it can iteratively learn from its mistakes.

\section{Conclusion}
Our exercises show that domain-specific LLM agents can streamline ground-based gamma-ray astronomy. By embedding project knowledge and enforcing validation, they generate reliable outputs that fit CTAO standards. A key lesson is a validation-first strategy: rather than trusting initial outputs, the agents iteratively check and refine them. Properly guided, modern LLMs can meet production requirements in our domain.

We also prioritize open-access LLMs. Using the Helmholtz Blablador platform\footnote{\url{https://strube1.pages.jsc.fz-juelich.de/2024-02-talk-lips-blablador/}}\footnote{\url{https://sdlaml.pages.jsc.fz-juelich.de/ai/guides/blablador_api_access/}}, we can swap between proprietary and self-hosted models for benchmarking and privacy-preserving, offline deployment.

In sum, these agents promise lower development overhead, fewer errors, and faster discovery in very-high-energy gamma-ray astronomy.
\section*{Acknowledgements}
We thank the Cherenkov Telescope Array Observatory, ACADA
Collaboration, and CTAO Medium-Size Telescope team for providing materials.

\bibliography{references.bib}

@ARTICLE{2025arXiv250300821K,
       author = {{Kostunin}, D. and {Sotnikov}, V. and {Golovachev}, S. and {Strube}, A.},
        title = "{AI Agents for Ground-Based Gamma Astronomy}",
      journal = {arXiv e-prints},
         year = 2025,
        month = mar,
          eid = {arXiv:2503.00821},
        pages = {arXiv:2503.00821},
          doi = {10.48550/arXiv.2503.00821},
archivePrefix = {arXiv},
       eprint = {2503.00821},
 primaryClass = {astro-ph.IM},
       adsurl = {https://ui.adsabs.harvard.edu/abs/2025arXiv250300821K}
}

@ARTICLE{2025arXiv250926110,
       author = {Kostunin, D. and Sotnikov, V. and Golovachev, S. and Mehta, A. and Holch, T.L. and Jones, E.},
        title = "{Agent-based code generation for the Gammapy framework}",
      journal = {arXiv e-prints},
         year = 2025,
        month = sep,
          eid = {arXiv:2509.26110},
        pages = {arXiv:2509.26110},
archivePrefix = {arXiv},
          doi = {10.48550/arXiv.2509.26110},
       eprint = {2509.26110},
 primaryClass = {cs.SE},
}

@article{Oya:2024dww,
    author = "Oya, I. and others",
    title = "{The first release of the Cherenkov Telescope Array Observatory array control and data acquisition software}",
    doi = "10.1117/12.3017568",
    journal = "Proc. SPIE Int. Soc. Opt. Eng.",
    volume = "13101",
    pages = "131011D",
    year = "2024"
}

@article{gammapy:2023,
author = {{Donath}, Axel and {Terrier}, R\'egis and {Remy}, Quentin and {Sinha}, Atreyee and {Nigro}, Cosimo and
{Pintore}, Fabio and {Kh\'elifi}, Bruno and {Olivera-Nieto}, Laura and {Ruiz}, Jose Enrique and
{Br\"ugge}, Kai and {Linhoff}, Maximilian and {Contreras}, Jose Luis and {Acero}, Fabio and
{Aguasca-Cabot}, Arnau and {Berge}, David and {Bhattacharjee}, Pooja and {Buchner}, Johannes and
{Boisson}, Catherine and {Carreto Fidalgo}, David and {Chen}, Andrew and {de Bony de Lavergne}, Mathieu and
{de Miranda Cardoso}, Jos\'e Vinicius and {Deil}, Christoph and {F\"u\ss{}ling}, Matthias and
{Funk}, Stefan and {Giunti}, Luca and {Hinton}, Jim and {Jouvin}, L\'ea and {King}, Johannes and
{Lefaucheur}, Julien and {Lemoine-Goumard}, Marianne and {Lenain}, Jean-Philippe and {L\'opez-Coto}, Rub\'en
and {Mohrmann}, Lars and {Morcuende}, Daniel and {Panny}, Sebastian and {Regeard}, Maxime and {Saha}, Lab
and {Siejkowski}, Hubert and {Siemiginowska}, Aneta and {Sip"ocz}, Brigitta M. and {Unbehaun}, Tim
and {van Eldik}, Christopher and {Vuillaume}, Thomas and {Zanin}, Roberta},
title = {{Gammapy: A Python package for gamma-ray astronomy}},
DOI= "10.1051/0004-6361/202346488",
url= "https://doi.org/10.1051/0004-6361/202346488",
journal = {A\&A},
year = 2023,
volume = 678,
pages = "A157",
}

@article{Kostunin:2025rlk,
    author = "Kostunin, Dmitriy and Jones, Elisa and Sotnikov, Vladimir and Sotnikov, Valery and Golovachev, Sergo and Strube, Alexandre",
    title = "{Enhancing the development of Cherenkov Telescope Array control software with Large Language Models}",
    doi = "10.22323/1.501.0714",
    journal = "PoS",
    volume = "ICRC2025",
    pages = "714",
    year = "2025"
}

\end{document}